\documentclass[letter]{aa} 

\usepackage{url}
\usepackage{graphicx}
\usepackage{natbib}
\usepackage[outercaption]{sidecap}
\def\SC@figure@vpos{t}
\bibpunct{(}{)}{;}{a}{}{,} 
\usepackage{txfonts}
\bibpunct{(}{)}{;}{a}{}{,} 

\def\vo{{\sc Virgo}}
\def\lo{{\sc LIGO}}
\def\ice{{\sc IceCube}}
\def\kmnet{{\sc KM3NeT}}

\begin{document}

   \title{Estimating source distances for high-energy neutrinos: A method for improving electromagnetic follow-up searches}
   \titlerunning{Estimating Source Distances for High Energy Neutrinos}

   \author{Thierry Pradier
          \inst{1}
          }

   \institute{Université de Strasbourg, CNRS, IPHC, UMR 7178, F-67000 Strasbourg, France\\
              \email{thierry.pradier@iphc.cnrs.fr}
         }

   \date{Received March 27th, 2023; accepted May 17th, 2023}

\abstract{High-energy neutrino telescopes such as \ice~or \kmnet~issue public alerts describing the characteristics of possible astrophysical high-energy neutrino events. This information, particularly with respect to the arrival direction and the associated uncertainty of the neutrino candidates, is used by observatories to search for possible electromagnetic counterparts. Such searches are complicated by the size of localisation areas, which can be up to tens of squared degrees or more, coupled with the absence of constraints on the distance or nature of the possible source -- in contrast to gravitational wave alerts issued by instruments such as \lo/\vo. Here, we describe method for deriving a probable distance interval for the astrophysical source that may possibly be associated with a HEN event, which may then be used in a cross-matching with galaxy catalogues to search for plausible electromagnetic counterparts. This study is intended to serve as a guide for high-energy neutrino followup campaigns.
}
   \keywords{high energy neutrinos   --
                Electromagnetic followup --
                multimessenger astronomy 
               }

   \maketitle



\section{Introduction}

High-energy neutrinos (HEN) are the smoking gun of the acceleration of hadronic cosmic rays in astrophysical sources \citep{halzen}, which are produced as a result of their interaction with radiation or matter. Since the discovery of a flux of cosmic HEN of TeV-PeV energies \citep{icecube2013}, a small number of objects have been clearly identified as possible HEN sources, namely: the blazar TXS0506+056, associated with the HEN IC170922A \citep{txsicecube},  the tidal disruption event (TDE) AT2019dsg \citep{tdestein} associated with IC191001A, and, more recently, the active galactic nucleus (AGN) NGC 1068 \citep{ngcicecube}. 
Other potential HEN sources range from short-duration transients, such as gamma-ray bursts (GRBs) \citep{grbwaxman}, with electromagnetic (EM) light curves fading after a few hours, to medium-duration transients, such as core-collapse supernovae (CCSN) \citep{ccsnmurase}, which end up fading over a few days or weeks. 
In long-duration transients, for instance, jetted TDE \citep{tdelitt}, the EM emission can last over a few months or years. The onset and duration of the HEN emission itself depends on the nature of the source.

Telescopes such as \ice~or \kmnet~\citep{km3netref} issue alerts after the identification of HEN candidates. 
In the case of \ice, these alerts are classified as 'gold' or 'bronze' depending on their estimated probability of being of astrophysical origin \citep{icalerts}. When such a HEN is identified, the arrival direction, plus an uncertainty on the order of 1$^{\circ}$ to 10$^{\circ}$ (depending on the neutrino type), together with a false-alarm rate and estimated energy, are broadcast in the form of a notice or circular. This information is then relayed to the astronomical community to look for potential transient EM counterparts to the HEN event and to help identify the origin of the detected HEN. This method is already in place for studies of gravitational wave (GW) alerts \citep{grandma} following \lo/\vo~notices.

With an observed rate of HEN alerts on the order of one per month for both channels, only a small fraction of these alerts are actually followed up on by public observation reports from observatories, despite the detection of new gamma-ray sources by FERMI-LAT in the 90\% uncertainty region (see e.g. \cite{ic220822A}) or the identification of possible counterparts by the Zwicky Transient Factory (see e.g. \cite{ztfexample}).
One of the reasons may be that the volume to be observed in the case of a HEN alert is unconstrained, which is not the case for GW alerts, for instance; for the latter,  3D information (i.e.  probable distance) is provided together with a 2D information (i.e. a probable position),  despite the greater localisation uncertainty for GW events (see e.g. \cite{gwexample}). Indeed the HEN error box generally covers an area well below the typical few hundreds of $\textrm{deg}^2$ for GW events \citep{lvlocalize}.
With the advent of the O4 observing run of the GW detectors \lo/\vo~to start before summer 2023 and its announced rate of one alert per day \citep{lvem}, it is of paramount importance to improve the information provided to electromagnetic observatories in order for them to efficiently look for potential transients associated with HEN candidates (see e.g. \cite{henztf}, \cite{asassnhen}). The aim of this paper is to show that such a probable distance interval can be derived for HEN sources in the case of a neutrino alert, since this information will facilitate the EM followups of HEN events. This is particularly relevant for short- and medium-duration transients for which a rapid response is critical because of their rapid fading.

The paper is organised as follows. Section~\ref{section1} describes the fundamental relationships between the energy and distance of HEN sources, which form the basis for our proposed method. Section~\ref{section2}  shows how the information dispatched by HEN telescopes such as \ice~can be used to constrain the search volume for electromagnetic followups, in particular, when they are combined with existing limits on the total energy emitted under the form of HEN. Section~\ref{section3} presents a possible practical implementation in the ranking of galaxies to be targeted in such searches for electromagnetic counterparts of HEN events.

\section{Number of neutrinos versus source properties}
\label{section1}

The number of HEN events detected in a neutrino telescope depends both on the source emission characteristics and the detector properties.

\subsection{HEN emission and detection} 

The total energy emitted in HEN during a transient emission, $E_{\textrm{iso}}^{\textrm{HEN}}$, as measured by the observer, for a source at a redshift, $z,$ and luminosity distance, $D_L$, is computed by integrating the neutrino spectrum over an energy range, $[E_{\textrm{min}}^{\textrm{obs}},E_{\textrm{min}}^{\textrm{obs}}]$, under the assumption that the source is emitting isotropically: 
\begin{equation}
\label{eq1}
\frac{E_{\textrm{iso}}^{\textrm{HEN}}}{4 \pi D_L(z)^2} = \int_{E_{\textrm{min}}^{\textrm{obs}}}^{E_{\textrm{max}}^{\textrm{obs}}} E \frac{dN}{dE} dE
.\end{equation}
\noindent
The corresponding minimum and maximum energies of the HEN emission are: $E_{\textrm{min, max}}^{\textrm{emit}} = E_{\textrm{min, max}}^{\textrm{obs}} \times (1+z)$ because of cosmic dilation. The HEN spectrum here in 1/GeV/cm$^2$ is generally expressed as a  power law  because of Fermi acceleration processes, which take the form: $\frac{dN}{dE} = \Phi_0 \left(\frac{E}{E_0}\right)^{-\gamma}$.

The average expected number of HEN events in a neutrino telescope $\langle N_{\textrm{HEN}}(\delta) \rangle$ can be estimated by convoluting the observed spectrum with $A_{\textrm{eff}}(\delta,E), $ namely, the energy and direction-dependent effective area for neutrinos of the telescope:
\begin{equation}
\label{eq2}
\langle N_{\textrm{HEN}}(\delta) \rangle = \int_{E_{\textrm{min}}^{\textrm{obs}}}^{E_{\textrm{max}}^{\textrm{obs}}} A_{\textrm{eff}}(\delta,E) \frac{dN}{dE} dE
.\end{equation}
\noindent
Here $E_{\textrm{min, max}}^{\textrm{obs}}$ stand for the energy range of the telescope. Using the published effective area for \ice~alerts \citep{icalerts}, this allows us, for instance, to compute for each alert the probability of observing at least one HEN event and the probability of observing exactly one HEN event depending on the direction of the alert:
\begin{align}
\label{eq3a} P_{\textrm{obs}}(N^{\textrm{obs}}_{\textrm{HEN}} = 1) &= \langle N_{\textrm{HEN}} \rangle e^{-\langle N_{\textrm{HEN}} \rangle} ,\\
\label{eq3b} P_{\textrm{obs}}(N^{\textrm{obs}}_{\textrm{HEN}} > 0) &= 1 - e^{-\langle N_{\textrm{HEN}} \rangle}.
\end{align}
\subsection{Relating the source energy and distance}

This average number $\langle N_{\textrm{HEN}} \rangle$ of expected HEN events can then be related to the total emitted energy in the form of HEN : 
\begin{equation}
\langle N_{\textrm{HEN}} \rangle = k_0(\gamma,\delta) \frac{E_{\textrm{iso}}^{\textrm{HEN}}}{4 \pi D_L^2}\label{eq4}
.\end{equation}
\noindent
The function $k_0(\gamma,\delta)$ depends on the assumed spectral index for the HEN emission and arrival direction especially, because of the direction-dependent effective area. This, in turn, allows us to estimate the probability of observing a number $N^{\textrm{obs}}_{\textrm{HEN}}$ of HEN events as a function of source energy and distance -- or, alternatively, the number of sources in the universe of a given energy at a given distance that can be detected with a number $N^{\textrm{obs}}_{\textrm{HEN}}$ of HEN events. The energy term at the numerator can be corrected for beaming effects with a factor of $\theta_{\textrm{jet}}^2$, with $\theta_{\textrm{jet}}$ as the HEN emission opening angle.

\section{Constraining the observation volume}
\label{section2}

Using the information provided, for instance, in \ice~notices issued after the detection of a candidate HEN, the most probable volume for finding a possible electromagnetic counterpart can be constrained by using the average number of HEN events expected as a function of energy and distance.

\subsection{Using the observation of a HEN candidate}

The first notice issued after the detection of a HEN track event (see \cite{1stnotice} for an example) yields the false-alarm rate (FAR) for the observed event. Given the total observation duration $T_{\textrm{obs}}$ (e.g. the time elapsed since the start of the alert system), the average expected number of events due to background only (i.e. atmospheric neutrinos) can be derived, so that an upper limit on $\langle N_{\textrm{HEN}} \rangle$ can be extracted if an astrophysical origin is assumed. With a minimum FAR of 0.15 event/yr for past \ice~alerts, the computed upper limit using Feldman-Cousins prescriptions is $\langle N_{\textrm{HEN}} \rangle_{\textrm{UL}} = 3.8$. For a maximum FAR of 4.9 events/yr as observed, the limit becomes $\langle N_{\textrm{HEN}} \rangle_{\textrm{UL}} = 0.9$.

\noindent
This translates into a condition on the distance of the source:
\begin{equation}
\label{eq5}
D_L \geq \sqrt{\frac{k_0(\gamma,\delta) \times E_{\textrm{iso}}^{\textrm{HEN}}}{4\pi \langle N_{\textrm{HEN}} \rangle_{\textrm{UL}}}}
.\end{equation}
\noindent
Figure \ref{fig_volume} (top) shows the distance lower limits as a function of the HEN energy of the source based on the observation of a single HEN event. Limiting the search for possible counterparts to realistic values of the HEN energy, marked by the colored bands (see for instance \cite{limitenergy}), already constrains (even if slightly), the volume of universe to be probed, by only targeting galaxies above the distance limit. Maximal HEN energies range from $1.5\times 10^{51}$ to $3\times 10^{52}$ erg for GRB-like sources and from $4 \times 10^{48}$ to $2 \times 10^{50}$ erg for supernova-like (SN-like) sources.

\subsection{Using the search for additional neutrinos}

A second notice (see \cite{2ndnotice} for an example) usually reports the search for additional muon neutrino events in the direction of the alert, in time intervals of $1000s$ and 2 days. The sensitivity $F_{\textrm{lim}}$ of the search for a given spectral index, equal to the upper limits in the case of no observation, is also provided together with the energy interval which would be valid for  $90\%$ of the events that \ice~would detect for this particular direction. This allows us to define a lower limit on the distance of the source: 
\begin{equation}
D_L \geq \sqrt{\frac{E_{\textrm{iso}}^{\textrm{HEN}}}{4\pi F_{\textrm{lim}} \times f(\gamma,E_{\textrm{min}},E_{\textrm{max}})}}\label{eq6}
,\end{equation}
\noindent
where $f(\gamma,E_{\textrm{min}},E_{\textrm{max}})$ depends on the spectral index, and where the information on the minimum and maximum energies of the search have to be considered. A beaming correction can also applied to the energy, depending on the type of sources envisaged.

\begin{figure}
\centering
\includegraphics[width=0.5\textwidth]{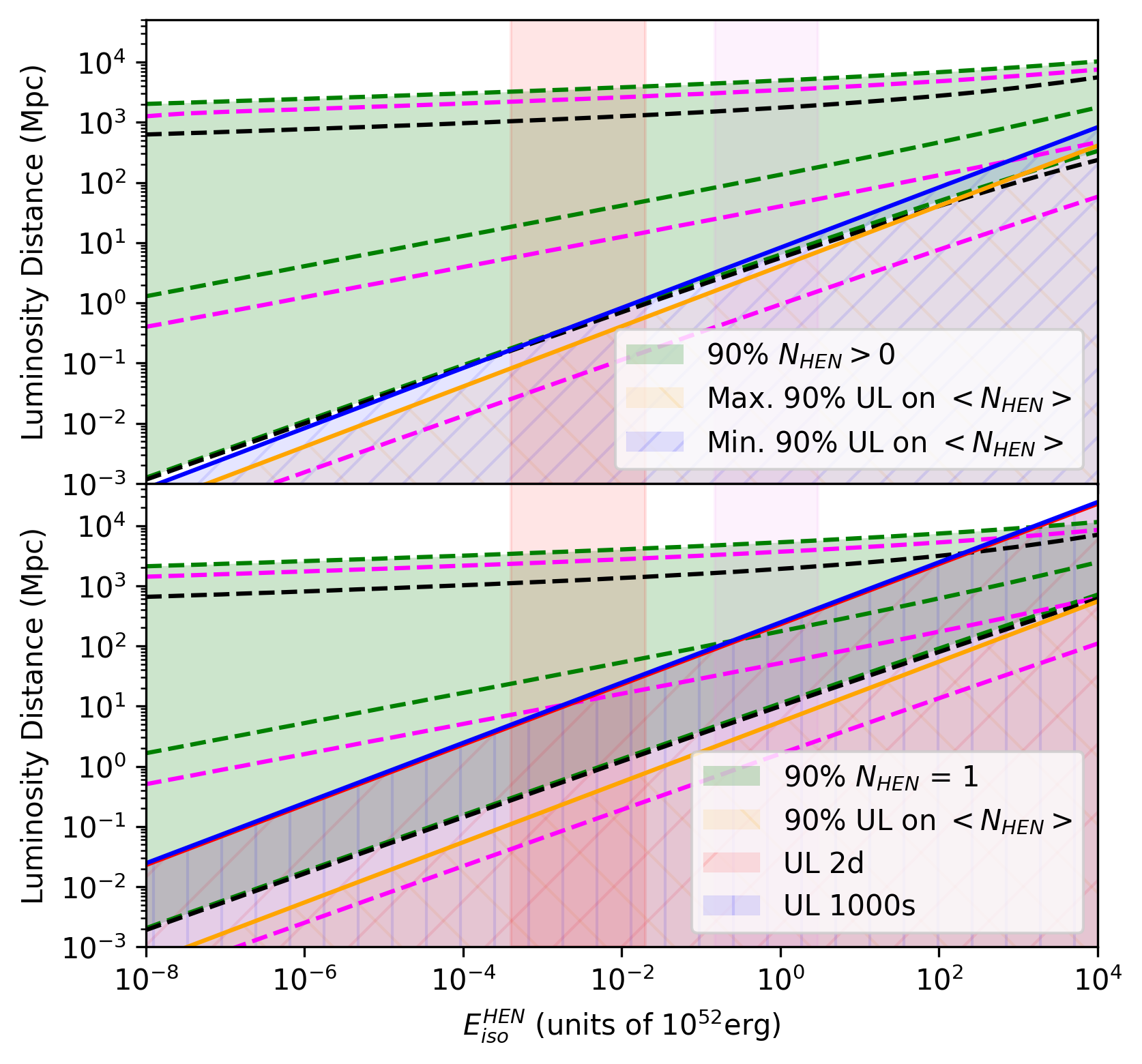}
\vskip -0.25cm
\caption{Distance of a source vs its isotropic HEN energy plan showing the probable 90\% interval  for the source distance, for different population and spectral indices. Top: after the first notice, when the number of observed HEN event is $1$, for a CCSN/SFR population assuming a spectral index $\gamma = 2$, in green. The yellow and blue region shows the region excluded after the first notice, given the minimum and maximum upper limits on the average number of HEN events. Bottom: Same quantities after the second notice reporting the search of additional HEN events and fluence limits in two different time windows (1000s and 2 days). In this plot, the limits are the ones given for the event \ice-230306A. The colored vertical bands show typical maximum energies for different classes of possible sources. The green dashed line indicate the median distances. The magenta lines show the modified 5\%-95\% interval for a GRB-population together with $\gamma = 2.5$, whereas the black lines refer to a population with no redshift evolution and $\gamma = 2$.}
\label{fig_volume}
\end{figure}

\noindent
Figure \ref{fig_volume} (bottom) shows the lower limit derived on the distance as a function of the HEN energy of the source, based on the upper limit on the HEN fluence. Limiting the search for possible counterparts to realistic values of the HEN energy further constrains the volume of universe to be probed -- again, by only targeting galaxies above the distance limit.

\subsection{Adding population information}

\begin{figure}
\centering
\centerline{\includegraphics[width=0.5\textwidth]{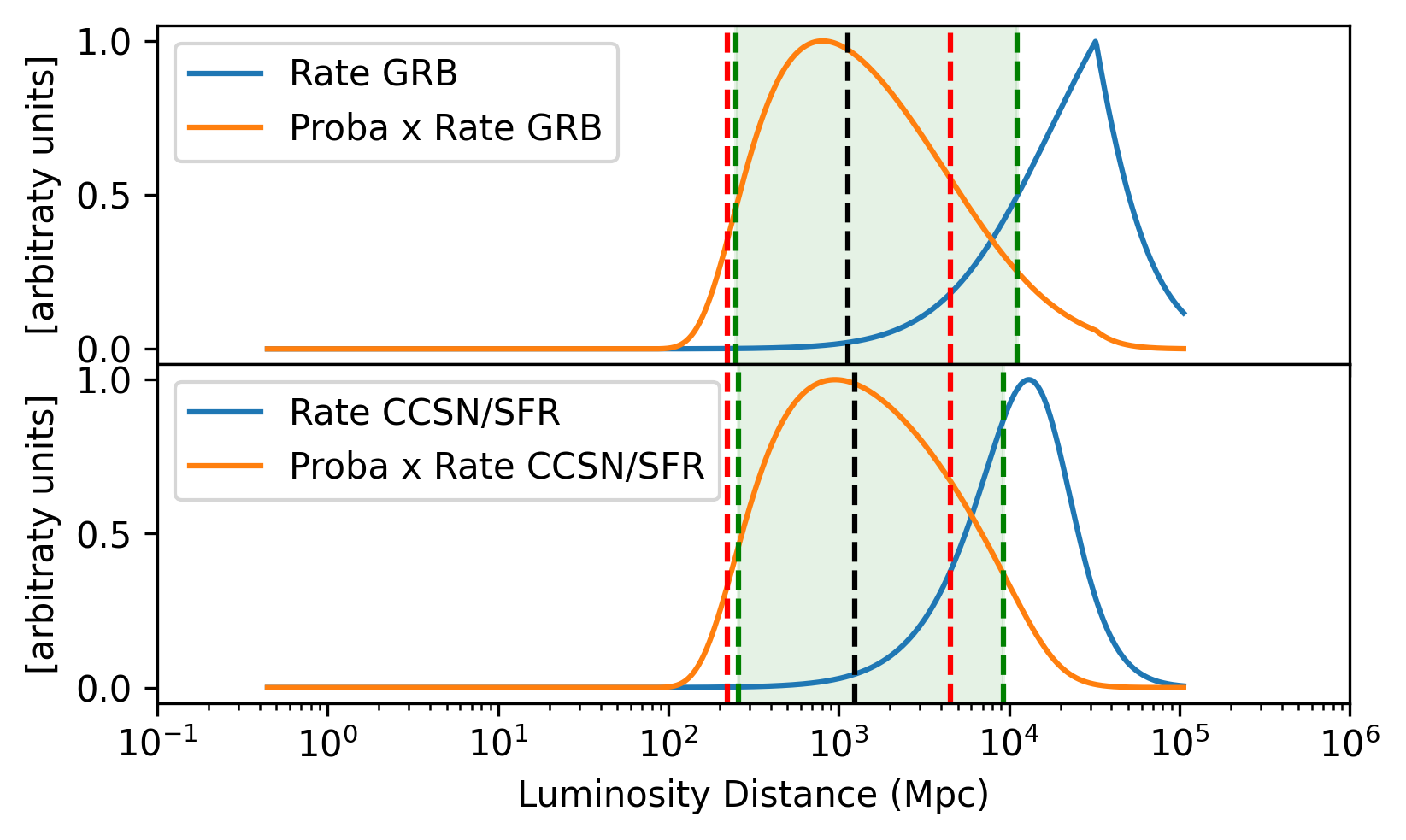}}
\vskip -0.25cm
\caption{Transient rates as a function of luminosity distance for CCSN and GRB populations and their convolution with $P_{\textrm{obs}}(N^{\textrm{obs}}_{\textrm{HEN}} = 1)$ for a given value of $E_{\textrm{iso}}^{\textrm{HEN}}$ (here $10^{55}$ erg for illustration purposes). The black dashed line indicate the median distance in each case, and the green band marks the 5\%-95\% percentiles, namely, the 90\% probable distance of the source. The red dashed lines represent the 5\%-95\% percentiles in the case of a transient rate with no redshift evolution.}
\label{fig_rate}
\end{figure}

The rate of transients per unit time is redshift-dependent :
\begin{equation}
\label{eq6}
R_{\textrm{transients}}(z) = \rho(z) \times \frac{dV_c}{dz}\times (1+z)^{-1},
\end{equation}
\noindent
where $\rho(z)$ is the transient density rate expressed in $1/\textrm{Mpc}^3/\textrm{yr}$, $dV_c/dz$ represents the differential comoving volume, and the additional redshift term accounts for cosmic time dilation. Here, we can, for instance, use typical CCSN redshift-dependent rates for $\rho(z),$
 which trace the star formation rate \citep{ccsnrate}, or GRB rates \citep{rategrb}, each with a different redshift dependence, or a source population with no redshift evolution. The many distant and faint sources will dominate over the fewer closer and brighter sources when it comes to detection. An example of rate evolution is shown in Figure \ref{fig_rate}.

\noindent
Convolved with the observation probability of $N^{\textrm{obs}}_{\textrm{HEN}} = 1$ event which depends on both distance $D_L$ and energy $E_{\textrm{iso}}^{\textrm{HEN}}$, a median distance and a 90\% interval for the distance can be defined, namely, the distance interval expected to contain 5\%-95\% of the sources with only one detected HEN. Finally, a 90\% containment volume can also be defined, once this distance constraint is combined with the localisation uncertainty of the particular HEN event, which is indicated in the alert notice. The resulting distance interval is shown in the two plots of Figure \ref{fig_volume}. It can be seen (top) that the sole information of the observation of a single HEN event is not enough to constrain the observation volume within the 90\% interval of the most probable distances. On the contrary, the absence of additional neutrinos and the fluence limit bring strong contraints on the most probable distance of the source (bottom), especially for isotropic energies above $10^{52}$ erg.

\noindent
The fraction of the 90\% distance of the source accessible within a given reach can be computed with a 50\% to 75\% probability for a source with $E_{\textrm{iso}}^{\textrm{HEN}} \le 3.3 \times 10^{49} \textrm{erg}$ detected with only one observed HEN event in \ice: it~must be located within 200~Mpc. This is an important guide for the organisation of searches for transient optical counterparts to HEN events.

\section{Practical implementation for a HEN followup}
\label{section3}

This information can be used to target galaxies falling inside the 90\% probable localisation region, given by the uncertainty on the arrival direction of a particular \ice~alert and within the derived 90\% distance range. This can be done thanks to a cross-match with a galaxy catalogue to extract the most probable galaxies of origin for the HEN event, for instance, the {\sc GLADE+} catalogue \citep{glade2p}. A ranking score can be defined for each galaxy of the catalogue for each $E_{\textrm{iso}}^{\textrm{HEN}}$, for instance, by: 
\begin{equation}
\label{eq7}
P_{\textrm{Galaxy}}(E_{\textrm{iso}}^{\textrm{HEN}}) = P_{\textrm{loc}}(\alpha_\textrm{Galaxy}, \delta_\textrm{Galaxy}) \times P_{\textrm{D}}(D_{\textrm{Galaxy}}),
\end{equation}
\noindent
where $P_{\textrm{loc}}(\alpha_\textrm{Galaxy}, \delta_\textrm{Galaxy}) \propto e^{-\frac{\Delta \theta^2}{2 \sigma^2}}$, in which $\Delta \theta$ represents the angular separation between the candidate galaxy and the HEN direction, and $\sigma$ a proxy for the angular resolution based on the 50\% and 90\% containment errors provided in the alert. Then, $P_{\textrm{D}}(D_{\textrm{Galaxy}})$ represents the resulting convolution of the transient rate and detection probability (presented in Figure \ref{fig_rate}). 

\subsection{Example of \ice-230306A}

Taking the example of \ice-230306A~\citep{lastic, lastic2}, which is labelled as a 'gold' event detected with an energy of 176 TeV on 6 March  2023, the probability density functions $P_{\textrm{loc}}$ and $P_{\textrm{D}}$ can be determined for a given energy, for instance, $E_{\textrm{iso}}^{\textrm{HEN}} = 3.3 \times 10^{49} \textrm{erg}$, 1\% of a typical GRB energy. The galaxy ranking combining the distance and localisation information is obviously different from the one that is based only on the localisation, as can be seen in Table \ref{tab_rank} (first and second columns). For $E_{\textrm{iso}}^{\textrm{HEN}} \in [10^{44}, 10^{51} \textrm{erg}]$, the selected galaxies are identical. With increasing values of $E_{\textrm{iso}}^{\textrm{HEN}}$, the selected galaxies are  increasingly distant. Given the ranges of energies expected for most HEN sources, it is safe to assume that $E_{\textrm{iso}}^{\textrm{HEN}} \lesssim 10^{52} \textrm{erg}$. 
For $E_{\textrm{iso}}^{\textrm{HEN}} =  10^{52} \textrm{erg}$, galaxies beyond $250$~Mpc must be discarded from the selection because of the fluence limit reported in the second notice.  When including this distance threshold in the cross-matching with the {\sc Glade+} catalogue, new galaxies are found. Their distances are shown in the fifth column of Table \ref{tab_rank} for $E_{\textrm{iso}}^{\textrm{HEN}} =  10^{52} \textrm{erg}$. For lower values of $E_{\textrm{iso}}^{\textrm{HEN}}$, the distance lower limit does not bring any constraint on the selection. On the other hand for $E_{\textrm{iso}}^{\textrm{HEN}} \gtrsim 10^{53} \textrm{erg}$, only a handful of galaxies are selected (last column of Table \ref{tab_rank}): the {\sc Glade+} catalogue is not sufficient, given the fluence limit at this energy yielding a distance lower limit of $\approx 800$ Mpc. When considering the different populations and spectrums, the selected galaxies do not change. 
\begin{table}
\caption{Distance and ranking, $R,$ of galaxies from the {\sc Glade+} catalogue for the cross-matching with HEN candidate IC230306A, using $P_{\textrm{D}}$ for $E_{\textrm{iso}}^{\textrm{HEN}} \in [10^{44} \textrm{erg}, 10^{52} \textrm{erg}]$ or without for $E_{\textrm{iso}}^{\textrm{HEN}} = 3.3 \times 10^{49} \textrm{erg}$. Only the first ten galaxies are shown. In bold we highlight those galaxies that are not ranked in the top five with no distance information. For $E_{\textrm{iso}}^{\textrm{HEN}} = 10^{52} \textrm{erg, we}$ show the galaxies found with or without taking into account the fluence limit provided by the \ice~notice. The symbol $(*)$ indicates the galaxies present in both the low- and high-energy scans. The average $\langle R \rangle$ uses $E_{\textrm{iso}}^{\textrm{HEN}} \in [10^{44}~\textrm{erg}-10^{52} \textrm{erg}]$ with distance information without a fluence limit. The last two columns include the fluence limit.}              
\label{tab_rank}      
\centering                                      
\begin{tabular}{c c c c | c c }          
\hline\hline                        
$D_L$ (Mpc)  & R & $10^{52}$ erg & $\langle R \rangle$ & $D_L$ (Mpc)  & $D_L$ (Mpc)\\ 
w. $P_{\textrm{D}}$ & no $P_{\textrm{D}}$ & no $F_{\textrm{lim}}$ & & $10^{52}$ erg  & $10^{53}$ erg \\
\hline                                   
 202.2  & 3             & 1  & 1 & 426.7 (*)    & 1208.6        \\
 150.0  & 1             & 2 & 2 & 701.4 (*)             & 842.4         \\
 106.7  & 2             & 6 & 4 & 418.9                 & 1080.4        \\      
 426.7 (*)      & {\bf 10}      & 3 & 3 & 306.0         & 861.3         \\
 140.7  & {\bf 6}       & 7 & 5 & 437.7                 & 859.4         \\
 170.9  & 7             & 5 & 5 & 386.8                 & 1131.9        \\
 234.3  & 9             & 4 & 7 & 307.0                 & -\\
 88.6   & 5             & 9 & 8 & 468.1                 & -\\
 701.4  (*) & >10       & >10 & >10 & 660.6             & - \\
 35.1   & 4                     & 9 & 9 & 567.2                 & -\\
\hline                                             
\end{tabular}
\end{table}

\subsection{Method for galaxy-targeted HEN followup}

Finally, we propose organising the search for EM transient counterparts of HEN events as follows: 
\begin{itemize}
\item[1-] At $T_0$ (first notice a few minutes after the detection of a HEN event),  the galaxies resulting from the cross-match with distance information are targeted, based on the average ranking obtained for $E_{\textrm{iso}}^{\textrm{HEN}} = 10^{44}~\textrm{erg} - 10^{52} \textrm{erg}$. This could save valuable time in the localisation of the potential counterpart.
\item[2-] At $T_1 \approx T_0 +1.5$ days (second notice reporting the fluence limit),  a low-energy ranking ($E_{\textrm{iso}}^{\textrm{HEN}} =  3.3 \times 10^{49} \textrm{erg}$) and a high energy ranking ($E_{\textrm{iso}}^{\textrm{HEN}} = 10^{52} \textrm{erg}$) are performed to target the selected galaxies in the search for associated EM counterparts.

\end{itemize}
\section{Conclusions}

Notices issued after the detection of a HEN candidate, currently by \ice~or soon by \kmnet~\citep{km3netalerts}, are used to constrain the volume of universe where a search for a possible associated EM counterpart can be carried out. 
Additional information, such as the nature of the source, the related source population, and its redshift-dependence, can also constrain the most probable distance of the HEN source.
Combined with maximal energies for the HEN emission obtained with previous searches by HEN telescopes, the most probable distance for the astrophysical source can be further constrained. The results of the cross-match with a galaxy catalogue are only slightly dependent on the assumed spectral index for the HEN emission or the source population.

As noticed in \cite{glade2p}, the {\sc Glade+} catalogue is complete up to $\approx 50$ Mpc in terms of the cumulative $B$-band luminosity of galaxies. For $E_{\textrm{iso}}^{\textrm{HEN}} \lesssim 10^{52} \textrm{erg}$, even if the probability for the astrophysical source to lie within 50 Mpc is only 30-70\% (depending on the assumed energy), the resulting probability for the source to originate from one of the cross-matched galaxies is still $30-70\%$. Moreover, the catalogue contains all of the brightest galaxies giving 90\% of the total $B$-band and $K_s$-band luminosity up to $\approx 130$ Mpc, where the probability of finding the astrophysical source is 40-75\%. The catalogue contains all of the brightest galaxies giving half of the total $B$-band ($K_s$-band) up to $\approx 250$ Mpc ($\approx 400$ Mpc), where the probability of finding the astrophysical source is 50-80\%. Finally, even if there is a 15-30\% probability that  the HEN source lies beyond 800 Mpc where the catalogue completeness falls below 20\%, it is still reasonable to look for a possible association with one of the closer catalogued galaxies. We recall that confirmed sources of HEN have been identified at distances between $\approx 10$~Mpc (NGC1068 \cite{ngcicecube}) and $\approx 1800$~Mpc (TXS0506+056 \cite{txsicecube}), whereas GW170817 at a distance of $\approx 40$~Mpc \citep{gw170817} was an envisaged HEN source \citep{gw170817hen}.

Combined with  long-term characterisations of existing high-energy gamma-ray sources, for instance, where blazars or AGNs could be present in the 90\% containment region of a given HEN alert (at the time of the HEN emission), such distance-constrained cross-matches could be useful in the pursuit of EM counterparts to HEN events, such as those reported in \cite{henztf} and \cite{asassnhen}.
\bibliographystyle{aa}
\bibliography{HENfollowup_searchvolume}

\begin{thebibliography}{29}
\expandafter\ifx\csname natexlab\endcsname\relax\def\natexlab#1{#1}\fi

\bibitem[{Abbasi {et~al.}(2018)}]{txsicecube}
Abbasi, R. {et~al.} 2018,
  \href{https://www.science.org/doi/abs/10.1126/science.aat1378}{Science}, 361,
  eaat1378

\bibitem[{Abbott {et~al.}(2020)}]{lvlocalize}
Abbott, B. {et~al.} 2020,
  \href{https://link.springer.com/article/10.1007/s41114-020-00026-9}{Living
  Reviews in Relativity}, 23, 3

\bibitem[{Abbott {et~al.}(2017)}]{gw170817}
Abbott, B.~P. {et~al.} 2017,
  \href{https://link.aps.org/doi/10.1103/PhysRevLett.119.161101}{Phys. Rev.
  Lett.}, 119, 161101

\bibitem[{Albert {et~al.}(2017)}]{gw170817hen}
Albert, A. {et~al.} 2017,
  \href{https://dx.doi.org/10.3847/2041-8213/aa9aed}{The Astrophysical Journal
  Letters}, 850, L35

\bibitem[{Antier {et~al.}(2020)}]{grandma}
Antier, S. {et~al.} 2020,
  \href{https://academic.oup.com/mnras/article/497/4/5518/5863231}{MNRAS}, 497
  Issue 4, 5518

\bibitem[{Assal {et~al.}(2021)}]{km3netalerts}
Assal, W. {et~al.} 2021,
  \href{https://dx.doi.org/10.1088/1748-0221/16/09/C09034}{Journal of
  Instrumentation}, 16, C09034

\bibitem[{D\'alya {et~al.}(2022)}]{glade2p}
D\'alya, G. {et~al.} 2022,
  \href{https://academic.oup.com/mnras/article/514/1/1403/6595338}{MNRAS},
  514-1, 1403–1411

\bibitem[{Halzen(2021)}]{halzen}
Halzen, F. 2021,
  \href{https://onlinelibrary.wiley.com/doi/abs/10.1002/andp.202100309}{Annalen
  der Physik}, 533, 2100309

\bibitem[{{IceCube}(2013)}]{icecube2013}
{IceCube}. 2013,
  \href{https://www.science.org/doi/10.1126/science.1242856}{Science}, 342,
  1242856

\bibitem[{{IceCube}(2022)}]{ngcicecube}
{IceCube}. 2022,
  \href{https://ui.adsabs.harvard.edu/link_gateway/2022Sci...378..538I/doi:10.1126/science.abg3395}{Science},
  378, 538

\bibitem[{{KM3NeT}(2019)}]{km3netref}
{KM3NeT}. 2019,
  \href{https://www.sciencedirect.com/science/article/pii/S0927650518302809}{Astroparticle
  Physics}, 111, 100

\bibitem[{{Madau} \& {Dickinson}(2014)}]{ccsnrate}
{Madau}, P. \& {Dickinson}, M. 2014,
  \href{https://www.annualreviews.org/doi/10.1146/annurev-astro-081811-125615}{Ann.
  Rev. of Astron. \& Astrophys.}, 52, 415

\bibitem[{Murase {et~al.}(2011)}]{ccsnmurase}
Murase, K. {et~al.} 2011,
  \href{https://link.aps.org/doi/10.1103/PhysRevD.84.043003}{Phys. Rev. D}, 84,
  043003

\bibitem[{Murase {et~al.}(2020)}]{tdelitt}
Murase, K. {et~al.} 2020,
  \href{https://dx.doi.org/10.3847/1538-4357/abb3c0}{Astrophys. J.}, 902, 108

\bibitem[{Necker {et~al.}(2022)}]{asassnhen}
Necker, J. {et~al.} 2022,
  \href{https://doi.org/10.1093/mnras/stac2261}{\mnras}, 516, 2455

\bibitem[{Reusch {et~al.}(2022)}]{tdestein}
Reusch, S. {et~al.} 2022,
  \href{https://ui.adsabs.harvard.edu/link_gateway/2022PhRvL.128v1101R/doi:10.1103/PhysRevLett.128.221101}{\prl},
  128, 221101

\bibitem[{{{\sc IceCube}}(2019{\natexlab{a}})}]{icalerts}
{{\sc IceCube}}. 2019{\natexlab{a}},
  \href{https://gcn.gsfc.nasa.gov/doc/IceCube_High_Energy_Neutrino_Track_Alerts_v2.pdf}{IceCube
  High-Energy Neutrino Track Alerts}

\bibitem[{{{\sc IceCube}}(2019{\natexlab{b}})}]{limitenergy}
{{\sc IceCube}}. 2019{\natexlab{b}},
  \href{https://link.aps.org/doi/10.1103/PhysRevLett.122.051102}{Phys. Rev.
  Lett.}, 122, 051102

\bibitem[{{{\sc IceCube}}(2023{\natexlab{a}})}]{1stnotice}
{{\sc IceCube}}. 2023{\natexlab{a}},
  \href{https://gcn.gsfc.nasa.gov/gcn3/33244.gcn3}{GCN Circular 33244
  IceCube-230201A}

\bibitem[{{{\sc IceCube}}(2023{\natexlab{b}})}]{2ndnotice}
{{\sc IceCube}}. 2023{\natexlab{b}},
  \href{https://gcn.gsfc.nasa.gov/gcn3/33256.gcn3}{GCN Circular 33256
  IceCube-230201A}

\bibitem[{{{\sc IceCube}}(2023{\natexlab{c}})}]{lastic}
{{\sc IceCube}}. 2023{\natexlab{c}},
  \href{https://gcn.gsfc.nasa.gov/gcn3/33403.gcn3}{GCN Circular 33403
  IceCube-230306A}

\bibitem[{{{\sc IceCube}}(2023{\natexlab{d}})}]{lastic2}
{{\sc IceCube}}. 2023{\natexlab{d}},
  \href{https://gcn.gsfc.nasa.gov/gcn3/33409.gcn3}{GCN Circular 33409
  IceCube-230306A}

\bibitem[{{{\sc IceCube}} {et~al.}(2021)}]{ztfexample}
{{\sc IceCube}} {et~al.} 2021,
  \href{https://heasarc.gsfc.nasa.gov/wsgi-scripts/tach/gcn_v2/tach.wsgi/?event=IC210629A&id=200164}{GCN
  Circulars related to IceCube-210629A}

\bibitem[{{{\sc IceCube}} {et~al.}(2022)}]{ic220822A}
{{\sc IceCube}} {et~al.} 2022,
  \href{https://heasarc.gsfc.nasa.gov/wsgi-scripts/tach/gcn_v2/tach.wsgi/?event=IC220822A}{GCN
  Circulars related to IceCube-220822A}

\bibitem[{{{\sc LIGO/Virgo}} {et~al.}(2021)}]{gwexample}
{{\sc LIGO/Virgo}} {et~al.} 2021,
  \href{https://gracedb.ligo.org/superevents/S200225q/view/}{Skymaps and GCN
  Circulars S200225q}

\bibitem[{{{\sc LIGO/Virgo/Kagra}}(2023)}]{lvem}
{{\sc LIGO/Virgo/Kagra}}. 2023,
  \href{https://dcc.ligo.org/LIGO-G2300151/public}{Low-latency update for Open
  LVEM}

\bibitem[{Stein {et~al.}(2023)}]{henztf}
Stein, R. {et~al.} 2023, \href{https://doi.org/10.1093/mnras/stad767}{\mnras},
  stad767

\bibitem[{{Wanderman} \& {Piran}(2010)}]{rategrb}
{Wanderman}, D. \& {Piran}, T. 2010,
  \href{https://academic.oup.com/mnras/article/406/3/1944/978260}{\mnras},
  406-3, 1944

\bibitem[{{Waxman} \& {Bahcall}(1997)}]{grbwaxman}
{Waxman}, E. \& {Bahcall}, J. 1997,
  \href{https://journals.aps.org/prl/abstract/10.1103/PhysRevLett.78.2292}{\prl},
  78, 2292

\end{thebibliography}

\end{document}